\documentclass{elsart3p}

\usepackage{natbib}

\usepackage{epsfig}
\usepackage{subfigure}
\usepackage{amsmath}

\usepackage{amssymb}

\def\mnras{MNRAS}
\def\apj{ApJ}
\def\apjs{ApJS}

\def\aap{A\&A}

\def\prd{Phys.~Rev.~D}

\begin{document}

\begin{frontmatter}



\title{Bayesian foreground analysis with CMB data}



\author[ita,cma,jpl,hke]{H. K. Eriksen}%
\author[caltech]{, C. Dickinson}%
\author[jpl]{, C. R. Lawrence}%
\author[sissa,infn,heidel]{, C. Baccigalupi}%
\author[mpa]{, A. J. Banday}
\author[jpl]{K. M. G\'{o}rski}%
\author[ita,cma]{, F. K. Hansen}%
\author[caltech2]{, E. Pierpaoli}%
\author[jpl]{ and M. D. Seiffert}


\address[ita]{Institute of Theoretical Astrophysics, University of
  Oslo, P.O. Box 1029 Blindern, 0315 Oslo, Norway}
\address[cma]{Centre of Mathematics for Applications, University of Oslo, P.O. Box
  1053 Blindern, 0316 Oslo, Norway}
\address[jpl]{Jet Propulsion laboratory, M/S 169-327, 4800 Oak Grove Drive,
  Pasadena, CA 91109}
\address[caltech]{Department of Astronomy, M/S 105-24, California Institute
  of Technology, 1200 E California Blvd, Pasadena, CA, 91125}
\address[sissa]{SISSA/ISAS, Astrophysics Sector, via Beirut 4, 34014
  Trieste}
\address[infn]{INFN, Sezione di Trieste, via Valerio 2, 34014 Trieste,
  Italy}
\address[heidel]{Institut f\"ur Theoretische Astrophysik, Universit\"at
  Heidelberg, Albert-Berle Strasse 2, 69120 Heidelberg, Germany}
\address[mpa]{Max-Planck-Institut f\"ur Astrophysik,
  Karl-Schwarzschildstrasse 1, Postfach 1317, 85741 Garching, Germany}
\address[caltech2]{Theoretical Astrophysics, M/S 130-33, California Institute
  of Technology, 1200 E California Blvd, Pasadena, CA, 91125}

\address[hke]{email: h.k.k.eriksen@astro.uio.no}

\begin{abstract}
  The quality of CMB observations has improved dramatically in the
  last few years, and will continue to do so in the coming decade.  Over a wide range of
angular scales, the uncertainty due to
  instrumental noise is now small compared to the cosmic variance.  One may claim
with some
  justification that we have entered the era of precision
  CMB cosmology.  However, some caution is still warranted: The errors
  due to residual foreground contamination in the CMB power spectrum
  and cosmological parameters remain largely unquantified, and the
  effect of these errors on important cosmological parameters such as
  the optical depth $\tau$ and spectral index $n_{\textrm{s}}$ is not
  obvious. A major goal for current CMB analysis efforts must therefore
  be to develop methods that allows us to propagate such uncertainties
  from the raw data through to the final products. Here we review a
  recently proposed method that may be a first step towards that goal.
\end{abstract}


\end{frontmatter}

\section{Introduction}
\label{sec:introduction}

The great importance of the foreground problem for CMB studies has
long been recognized in the cosmological community, and as a result a
large number of algorithms have been proposed, implemented and applied
to both simulated and real data. Examples are the Maximum Entropy
Method
\citep{barreiro:2004,bennett:2003b,hobson:1998,stolyarov:2002,stolyarov:2005},
Internal Linear Combination methods
\citep{bennett:2003b,tegmark:2003,eriksen:2004a,saha:2005,hansen:2006,hinshaw:2006},
Wiener filtering \citep{bouchet:1999,tegmark:1996}, Independent
Component Analysis methods \citep{maino:2002,maino:2003,donzelli:2005}
and spectral fitting approaches \citep{delabrouille:2003}.

The main emphasis of most of these methods lies on establishing an
optimal estimate of the CMB sky signal, and not so much on quantifying
the uncertainty on that estimate. A different approach was 
taken by \citet{brandt:1994} who simply adopted well-established
likelihood parameter estimation techniques to solve the problem. Ten
years later this work was continued by \citet{eriksen:2006}, by taking
advantage of recent statistical (in particular Markov Chain Monte
Carlo) and computational developments. 

In this paper we briefly review the algorithm as presented by
\citet{eriksen:2006}. We then describe three applications, namely 1)
an analysis of simulated Planck + six-year WMAP data, 2) a preliminary
analysis of the first-year WMAP data, and 3) a preliminary study of
experimental design.

\section{Methods}
\label{sec:algorithms}

The most straightforward method for producing reliable error bars is
provided by standard parameter estimation techniques. In particular,
the Bayesian framework is particularly suited for this type of
problem.

Given a set of observed data $\mathbf{d}$ and an assumed model
$\mathbf{s}(\theta)$, $\theta$ being a general set of parameters, we
simply ask, what is the posterior distribution $P(\theta|\mathbf{d})$?
To answer this, we first recall Bayes' formula,
\begin{equation}
P(\theta|\mathbf{d}) \propto P(\mathbf{d}|\theta) P(\theta) = \mathcal{L}(\theta)P(\theta),
\end{equation}
where $\mathcal{L}(\theta) \equiv P(\mathbf{d}|\theta)$ is the
likelihood and $P(\theta)$ is a prior summarizing our previous
knowledge about $\theta$. For Gaussian data, the likelihood is given
by 
\begin{equation}
-2\ln\mathcal{L} = \chi^2(\theta) = \left[\mathbf{d} -
  \mathbf{s}(\theta)\right]^t \mathbf{C}^{-1} \left[\mathbf{d} -
  \mathbf{s}(\theta)\right],
\end{equation}
where $\mathbf{C}$ is the covariance matrix, up to an irrelevant
constant. (In the current implementation, we assume no pixel-pixel
correlations, and $\mathbf{C} = \mathbf{N}$ is defined to be the noise
covariance matrix.) The problem is thus reduced to mapping out the
posterior as a function of the free parameters $\theta$ using some
computational tool, such as MCMC or grid evaluations.

\begin{figure*}[t]
\mbox{\epsfig{figure=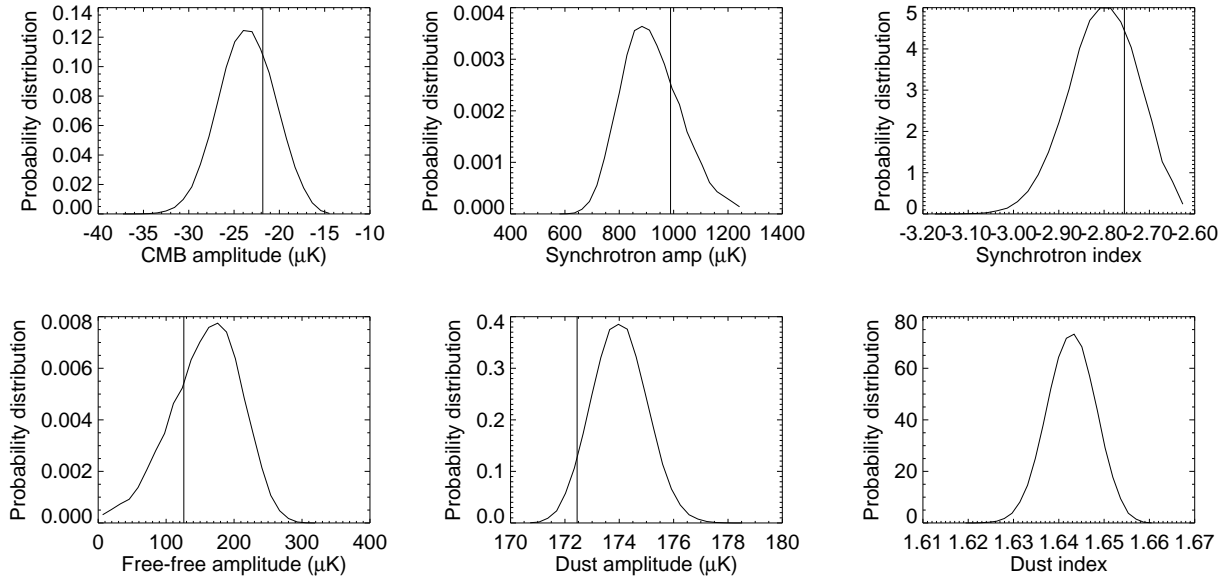,width=\linewidth,clip=}}
\caption{Marginalized posterior distributions for six parameters
  fitted to a single pixel by Markov Chain Monte Carlo. This example
  is taken from a simulation corresponding to the combination of
  Planck and six-year WMAP, for a pixel located exactly on the
  Galactic plane. The vertical lines indicate the true parameter
  values. (Note that this is not a well-defined quantity for the dust
  spectral index, since we fit for a one-component dust model whereas
  the simulation is based on a two-component dust model.)}
\label{fig:onedim_dist}
\end{figure*}

This machinery may be applied to microwave component reconstruction by
noting that different signal components have different spectral and
spatial behaviour. By observing the sky in different frequencies and
directions one may 
disentangle the
various contributions and
isolate the CMB
signal.

The first step is therefore to choose a suitable parametric model. For
CMB analysis, a particularly convenient choice is that of a sum of
independent modulated power-law components,
\begin{align}
S(\nu; \theta) &= \sum_{i=1}^{N_{\textrm{comp}}} S_{i}(\nu; \theta_i)
\\ &=
\sum_{i=1}^{N_{\textrm{comp}}} A_{i} f_i(\nu)
\left(\frac{\nu}{\nu_{0,i}}\right)^{\beta_{i}}.
\end{align}
Here one typically would include a CMB term ($A_{\textrm{cmb}} =
T_{\textrm{cmb}}$; $f_{\textrm{cmb}}(\nu)$ is the
thermodynamic-to-antenna temperature conversion factor;
$\beta_{\textrm{cmb}} \equiv 0$), a synchrotron and free-free term
($A$ and/or $\beta$ free parameters; $f_{\textrm{s}}(\nu) = 1$), and a
thermal dust term.

These few simple definitions summarize the method quite succinctly.  No a priori assumptions
about the sky emission are required, except that it can be represented by a parametric model.
The remaining discussion is concerned mostly about how to deal with real-world complications
such as limited spectral information, low signal-to-noise ratio, and computational
constraints.  It is worth emphasizing that even an unrealistically simple sky model
containing nothing but CMB, free-free,  synchrotron, and dust emission requires at least
six parameters, and that no CMB experiment to date has had even the eight frequencies
required to fit such a minimal model.  Fortunately, the method can be extended easily to
include other information, as will be seen in the next section.

\subsection{Calibration and template fitting}

Two major complications arise when trying to apply the above machinery
to real-world data. First, spectral fits generally require that all
sky maps are properly calibrated with respect to a common zero-point.
This is not trivial for any CMB experiment, and certainly not for
differential observatories such as the WMAP satellite. Second, as previously mentioned, the
number of observed frequencies is often (i.e., in every experiment performed to date!) smaller
than the number of parameters one might wish to include.  For instance, WMAP observes the
sky at five frequencies, while, ideally, we would like to include at
least six parameters in a reasonable model (CMB, synchrotron,
free-free and thermal dust amplitudes, and synchrotron and dust
spectral indicies), still neglecting a possible anomalous dust
contribution.

A straightforward way to address both problems is to
introduce template terms in the signal model,
\begin{equation}
\begin{split}
S(\nu; \theta) \rightarrow S(\hat{n}, \nu; \theta) =\quad\quad\quad\quad\quad\quad\quad \\
= \sum_{i=1}^{N_{\textrm{pix}}} S(\nu;\theta_i) +
\sum_{i=1}^{N_{\textrm{temp}}} A_i(\nu) g_i(\hat{n}).
\end{split}
\end{equation}
That is, in addition to the previously defined (and usually rather
complicated) spectral models, we also fit for a small set of global
(and simple) template amplitudes. Common templates to include would be
a monopole term, three dipole terms, and either reasonably
well-behaved (e.g,. free-free emission through an H$_\alpha$ template)
or sub-dominant (e.g., thermal dust emission for WMAP) foreground
components.  For each signal component one is able to handle this way,
one saves one degree of freedom in all subsequent non-linear fits,
thus obtaining very useful additional stability.

The cost of this approach is that the problem becomes
global, and the computational load is vastly increased.  At present it
is therefore only feasible to perform such an analysis at low
resolution and with a non-linear search.  Nevertheless, since the main
target at this stage is a relatively small number of global template
amplitudes, little is lost by this approach.

Note that this extension of the method was discussed only briefly by
\citet{eriksen:2006} in the context of monopole and dipole
calibration, not general template fits, and it was also not
implemented at that time. The analysis of the first-year WMAP data
presented in this paper is therefore the first application of the
method.

\subsection{Estimating spectral indices from low-resolution maps}


After calibrating and removing well-behaved components from the maps
using the above description, the next step is to map out the full
posterior for each pixel with low-resolution maps. The reasons for
using low-resolution maps at this stage are two-fold: On the one hand,
estimation of non-linear parameters such as spectral indices is quite
sensitive to noise, and it is therefore highly desirable to have high
signal-to-noise data. On the other hand, we use full-blown MCMC to map
out the posteriors for each pixel, and with a computational cost of
about 100 CPU-seconds per pixel it is currently not feasible to do
this for mega-million pixel maps.

\begin{figure}[t]
\mbox{\epsfig{figure=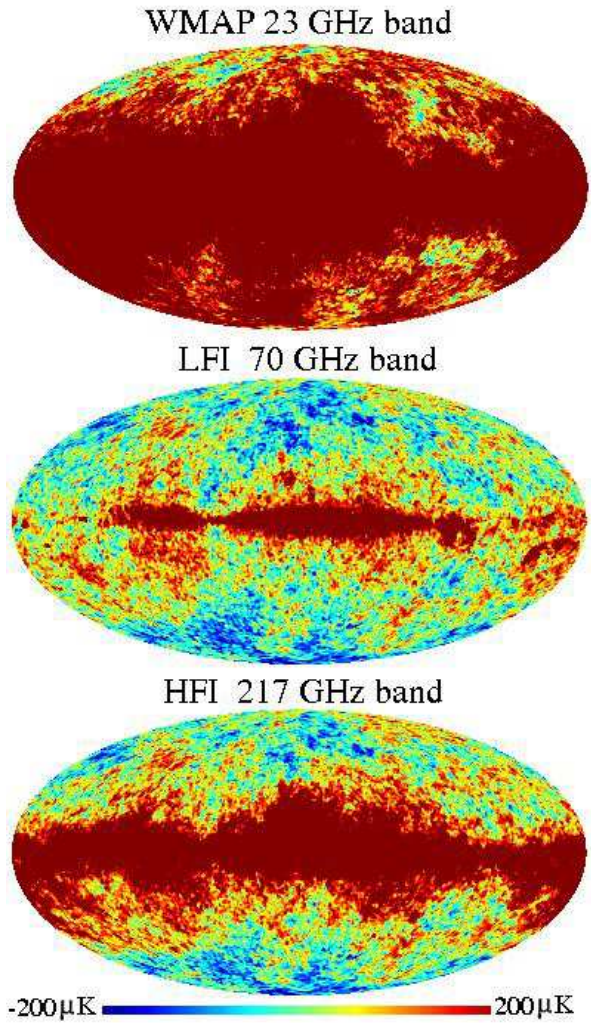,width=\linewidth,clip=}}
\caption{Simulation constructed to test the algorithm based on a
  semi-realistic model of the foregrounds: The synchrotron emission
  includes spatial variations in the spectral index
  \citep{giardino:2002}, and the thermal dust emission follows the
  two-component ``model 8'' of \citet{finkbeiner:1999}. For further
  information on this simulation, see \citet{eriksen:2006}.}
\label{fig:simulation}
\end{figure}

For these two reasons, we smooth each map with a wide beam (typically
with a Gaussian beam of, say, $7^{\circ}$ FWHM) and downgrade the
pixel resolution (typically to, say, $2^{\circ}$ pixel size;
$N_{\textrm{side}} = 32$ in HEALPix
language\footnote{http://healpix.jpl.nasa.gov; \citet{gorski:2005}})
prior to the MCMC analysis. The computational expense is then quite
manageable with a modern-type cluster, with a total cost of about 500
CPU hours.

The final product from this stage is a single joint probability
distribution for each pixel, and from these we find the univariate
distributions for each parameter by marginalizing over all others.
Examples of such univariate distributions are shown in Figure
\ref{fig:onedim_dist}. Finally, we store the posterior mean and
variance for each parameter and each pixel in the form of two sky
maps as our final data products.

\subsection{Estimating amplitudes from high-resolution maps}

\begin{figure*}[t]
\mbox{\epsfig{figure=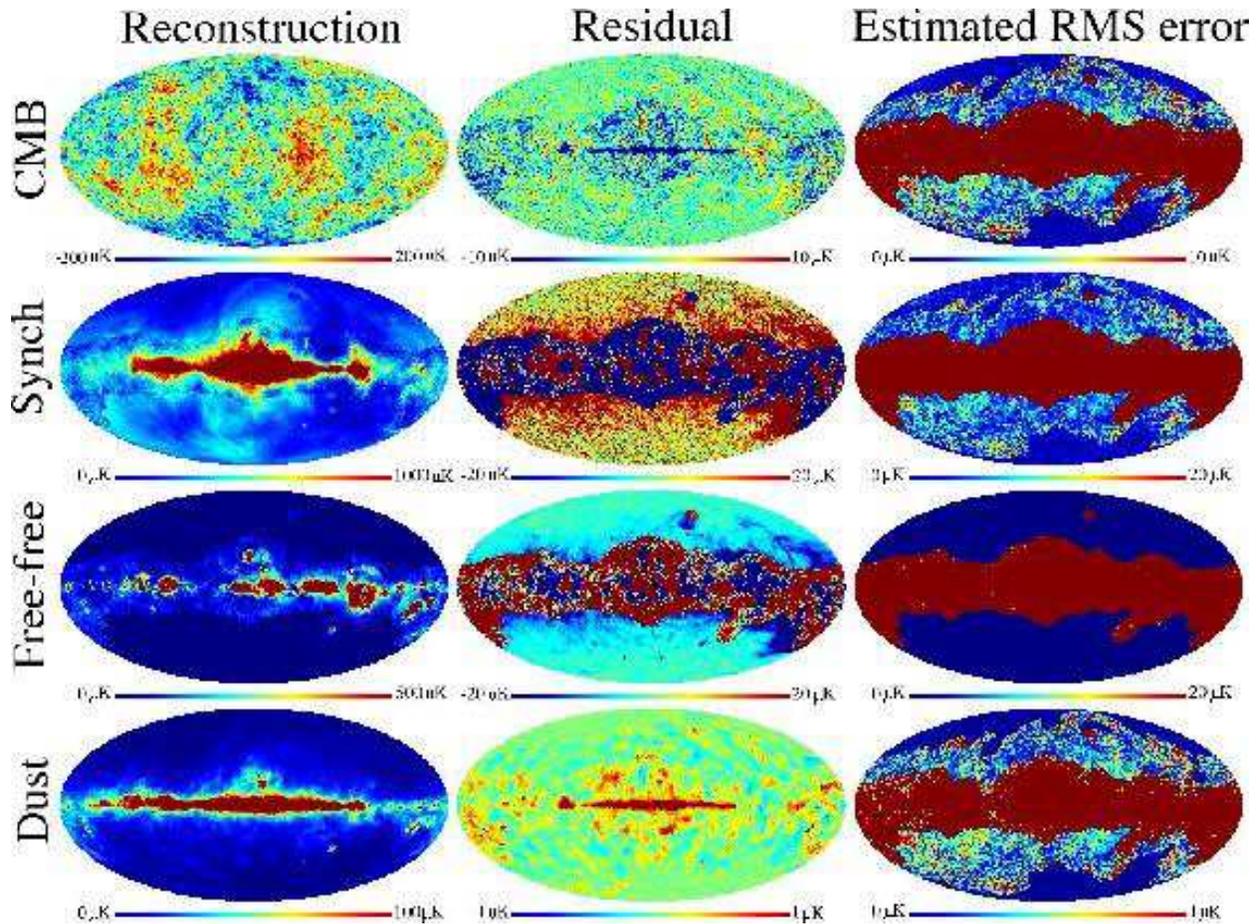,width=\linewidth,clip=}}
\caption{Full-sky results obtained from the simulation described in
  Section \ref{sec:simulation}. Shown are reconstructed parameter
  values (left column), actual residual errors (middle column), and
  estimated uncertainties (right column). The sharp boundaries are
  caused by different assumed models (e.g., free-free is omitted from
  the fitting model at high latitudes but included at low), which
  leads to order-of-magnitude different uncertainties because of
  degeneracies.}
\label{fig:sim_fullres}
\end{figure*}

While the procedure described in the previous section gives a full
representation of the posteriors for the low-resolution sky maps, we
also want a good representation of the full-resolution posteriors. To
obtain these, we make the assumption that the spectral indices vary
more slowly than the amplitudes\footnote{Whether this assumption is
  well justified will only be clear after the first
  high-quality observations (from Planck) are analyzed, but based on the limited data
available to date, this looks to be a reasonable approximation.}, and fix their
distributions at the corresponding low-resolution values. \emph{Given} the values of all
non-linear parameters, the likelihood becomes Gaussian, and the
analysis may be performed analytically with modest computational
expenses. For details on this procedure, see \citet{eriksen:2006}.

\section{Example applications}
\label{sec:applications}

In the following sections, we show a few demonstrations of how this
method performs in practice.  First, we review the simulation
described by \citet{eriksen:2006}.  We then show the
results from a preliminary analysis of the first-year WMAP data.
We emphasize that these results are preliminary, and will be
revised with the currently available three-year data.  Finally, we show
some early results from a currently on-going study of optimization of
future experiments.

\subsection{Planck + six-year WMAP simulation}
\label{sec:simulation}

In order to test the algorithm, \citet{eriksen:2006} constructed a
simulation corresponding to a combination of six Planck channels (30
to 217 GHz) and five WMAP channels (23 to 94 GHz). The
higher-frequency Planck channels were not included because of
modelling error confusion associated with including multiple dust
components. The simulation took into account the beam and white noise
characteristics of each channel separately. Four signal components
were included: a Gaussian CMB realization drawn from the best-fit WMAP
$\Lambda$CDM power spectrum; a semi-realistic synchrotron component,
featuring a spatially varying spectral index \citep{giardino:2002}; a
free-free component with fixed spectral index of $\beta_{\textrm{ff}}
= -2.15$ \citep{dickinson:2003}; and a two-component thermal dust
model (``model 8'' of Finkbeiner et al.\ 1999). Thus, the simulation
includes two major complications that we must expect to find in real
data, namely both spatially varying spectral indices and departures
from simple power law spectra. Three selected channel maps are shown
in Figure \ref{fig:simulation}.

These simulations were then analyzed using the machinery described
above.  The results are discussed in detail in the original
paper; two sample results are reprinted here. In Figure
\ref{fig:onedim_dist} we show the marginal densities for one single
pixel located exactly on the Galactic plane, and the true values for
each parameter is marked by a solid line. Clearly, the posterior
distributions agree very well with the input values, both in mean and
variance. 

In Figure \ref{fig:sim_fullres} we show sky maps of the four component
amplitudes that were estimated: Posterior means are shown in the left
column, actual residuals are shown in the middle column, and estimated
errors are shown in the right column. Again, the results agree very
well with expectations.

\subsection{The first-year WMAP data}
\label{sec:wmap}

In this section we present for the first time the results from a
parametric foreground analysis of the first-year WMAP data
\citep{bennett:2003a}. These data comprise a set of full-sky
temperature maps at five frequencies (23, 33, 41, 61 and 94 GHz) at
angular resolutions between $13'$ and $53'$ FWHM. The noise is assumed
to be white, and all beams are assumed to be circularly symmetric.

As described above, the analysis was performed in three stages, two at
low resolution ($7^{\circ}$ FWHM for WMAP), and the third at high
resolution ($1^{\circ}$). In the first or calibration stage, the free
parameters were a monopole and dipole amplitude, a free-free template
amplitude \citep{finkbeiner:2003}, and a dust template amplitude
\citep{finkbeiner:1999} for each band, and a CMB temperature and a
synchrotron amplitude and spectral index for each pixel. The fit was
performed using a non-linear Sequential Quadratic Programming (SQP)
algorithm.

\begin{figure}[t]
\centering
\mbox{\epsfig{figure=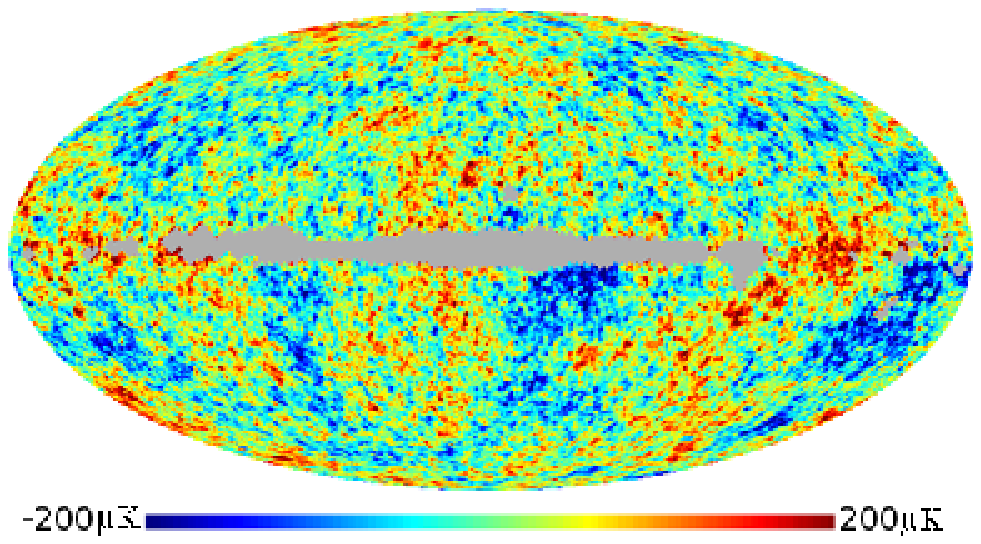,width=\linewidth,clip=}}
\mbox{\epsfig{figure=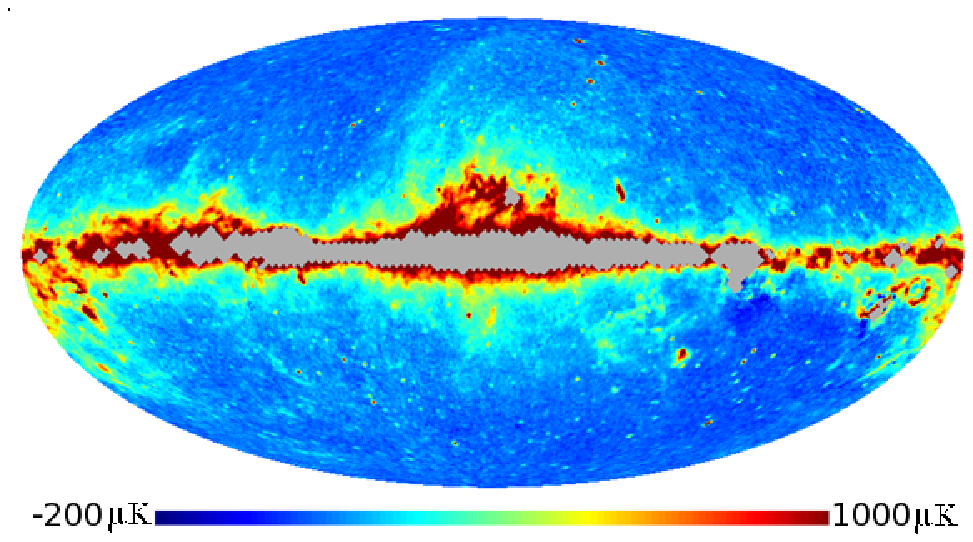,width=\linewidth,clip=}}
\mbox{\epsfig{figure=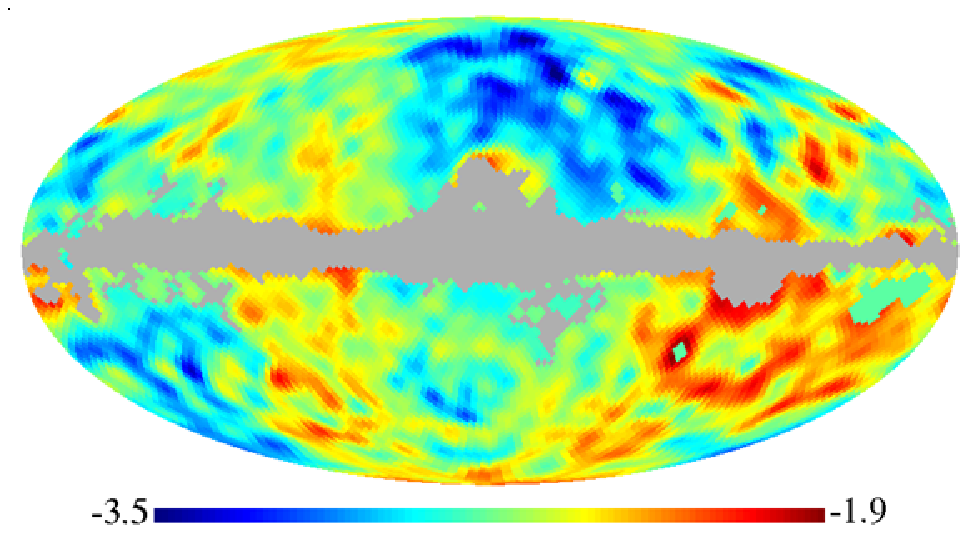,width=\linewidth,clip=}}
\caption{Results from first-year WMAP analysis. Shown are
  reconstructed CMB signal (top panel), synchrotron amplitude (middle
  panel), and synchrotron spectral index (bottom panel).}
\label{fig:wmap1}
\end{figure}

In the second stage, the low-resolution posterior distributions are
mapped out using \hbox{MCMC}. Only CMB and synchrotron (both amplitude
and spectral index) were included in this case. The main result from
these computations is a synchrotron spectral index map, shown in the
bottom panel of Figure~\ref{fig:wmap1}. Note that the most values lie
between $-2.0$ and $-3.4$, which is quite acceptable. (Values larger
than $-2.3$ probably indicate residual free-free emission, rather than
break-down in the synchrotron estimation.)

In the third stage, we estimate the CMB and synchrotron amplitudes
analytically.  The results from these calculations are shown in the
two panels of Figure~\ref{fig:wmap1}. (The synchrotron amplitude is
normalized to K band.) Again, the results appear visually quite
compelling, although a more quantitative analysis is warranted.
However, we re-emphasize that these results are presented only to
demonstrate the capabilities of the method, and not as definitive
measurements. The analysis will soon be revisited based on the
recently available three-year WMAP data.

\subsection{Experiment design}
\label{sec:expdesign}

The machinery described in Section \ref{sec:algorithms} is also very
well suited to study optimization of future experiments. The question
we want to answer is, given a set of instrumental constraints (e.g.,
focal plane area and power dissipation) how should we distribute the
number of detectors as a function of frequency in order to optimize
our ability to extract the CMB signal? 

One possible method for answering this question would be to apply the
above machinery to a wide range of allowed instrument configurations,
and study how the uncertainties (or residuals) change with
configuration. In a currently on-going project, we try to do this in a
systematic fashion, and in this section we present some very
preliminary results from this study.

\begin{figure}[t]
\mbox{\epsfig{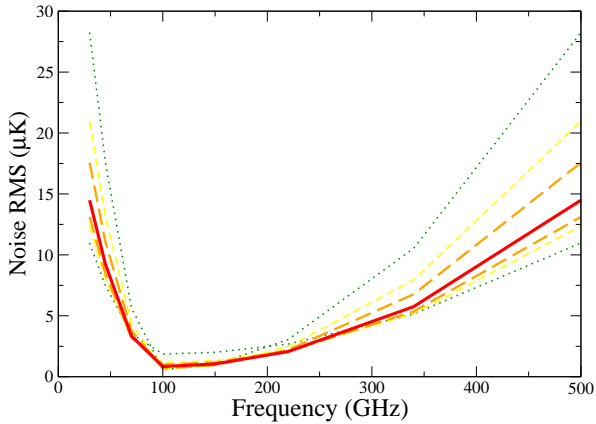}}
\caption{Example results from the experimental design study. The
  curves represent different configurations allowed by the
  experimental constraints, and show the effective noise per frequency
  channel, which is directly related to the number of detectors per
  frequency.  The thickness (and color) indicates the reconstruction
  quality: A thicker (and redder) line indicates a better
  reconstruction. The optimal solution here is the thick red line,
  which corresponds to a constant signal-to-noise ratio.}
\label{fig:expdesign}
\end{figure}

We consider a generalized version of the simulations described in
Section~\ref{sec:simulation}, for which the frequency bands and the noise RMS per frequency
may be chosen freely. We then impose a set of constraints corresponding to the focal plane
area of an optical system that could fly in space, and feeds at different frequencies of
realistic size. Based on these constraints, we generate a grid of possible instrument
configurations, and run the analysis for a reasonable set of these. (Most may be discarded by
common sense.) 

A few example results are shown in Figure \ref{fig:expdesign}: The
lines indicate the noise RMS distribution as a function of frequency
($\sigma(\nu) = \sigma_0(\nu) /\sqrt{N_{\textrm{det}}(\nu)}$, where
$N_{\textrm{det}}$ is the number of detectors at frequency $\nu$) for
seven different configurations. The thickness (and color) of the lines
indicates the resulting uncertainty/residual from the component
separation: A thicker (redder) line means a better reconstruction. The
optimal distribution of detectors among this set is therefore the
solid red line.

From this simple exercise, we may formulate a first design principle:
\emph{The optimal noise distribution is the one that corresponds to a
  constant signal-to-noise ratio, where the signal is the sum of CMB
  and foregrounds.} Second, the wider the frequency coverage, the
better the reconstruction.

However, there is a major caveat here: This principle only holds when
any modelling errors (i.e., errors due to the fact that the assumed
parametric model is different from the true signal) are small.
Performing a similar exercise on simulations with significant
modelling errors results in a competing principle: \emph{When the
  parametric shape of the foreground components are poorly known, the
  best reconstructions are obtained with configurations surrounding
  the foreground minimum.} The optimal total frequency span depends on
the magnitude of the modelling errors. Both of these principles sound
rather obvious, but they are nevertheless worth making explicit.

In conclusion, two competing effects are at work: Reconstruction power
is maximized by a wide frequency coverage, but limited by modelling
errors. In a future publication, we will quantify these considerations
in much greater detail, and attempt to provide some guidelines on the
preferred frequency ranges for future CMB experiments.  
Firm conclusions, however, will require an understanding of modelling errors based on better
knowledge of the foregrounds themselves, knowledge that can come only from measurements of
the polarized sky over a broad range of frequencies with great sensitivity.

\section{Discussion}
\label{sec:discussion}

As the sensitivity of CMB observations continue to improve, the
importance of properly characterizing the foreground contributions
also increases. Even with WMAP we are already at a level where
instrumental noise is negligible compared to the foreground
uncertainties over a wide range of angular scales
\citep{hinshaw:2006}, and this will be even more true for Planck and
up-coming polarization experiments.
Simple template corrections will be hopelessly inadequate 
to reach the required sensitivity levels.  Further, it will be
essential to propagate the foreground uncertainties through to the
final products, namely the CMB power spectrum and cosmological
parameters.

On this background, we argue that the most appropriate solution will
be based on traditional parameter estimation techniques, rather than
image processing techniques. The reason is simply that such methods
naturally provide the full posterior
distributions, and are integrated easily with existing power spectrum and parameter
estimation methods based on Bayesian parameter estimation.  An approach that
appears particularly promising in this respect is that of Gibbs sampling,
which allows for joint global analysis of foregrounds, CMB power
spectrum and even cosmological parameters \citep{jewell:2004,
  wandelt:2004, eriksen:2004b, eriksen:2006}.

In fact, we end by emphasizing that the particular implementation we
describe in this paper is of less importance than its underlying idea.
Many improvements can be made to the algorithm as such (e.g.,
introducing support for spatial correlation information would be of
great value), but our main conclusion is independent of such details:
CMB component separation is a probabilistic problem, and obtaining
accurate uncertainties is a crucial part of the problem. Parameter
estimation techniques provide the most direct route for doing so.

\end{document}